\documentclass{PoS}

\def \apj {ApJ}
\def \apjl {ApJL}
\def \apjs {ApJS}
\def \aap {A\&A}

\def \aj {Astronomical Journal}
\def \mnras {MNRAS}

\title{The role of structured OB supergiant winds in producing the X-ray flaring emission from High Mass X-ray Binaries}

\ShortTitle{Accretion of clumpy wind in supergiant HMXBs}

\author{\speaker{L.~Ducci}$^{ab}$,
  L.~Sidoli$^b$, A.~Paizis$^b$, S.~Mereghetti$^b$, P.~Romano$^c$\\
  \llap{$^a$} Dipartimento di Fisica e Matematica, Universit\`a degli
  Studi dell'Insubria,\\ 
  Via Valleggio 11, I-22100 Como, Italy \\
  \llap{$^b$} INAF, Istituto di Astrofisica Spaziale e Fisica
  Cosmica,\\
  Via E. Bassini 15, I-20133 Milano, Italy \\
  \llap{$^c$} INAF, Istituto di Astrofisica Spaziale e Fisica
  Cosmica,\\
  Via U. La Malfa 153, I-90146 Palermo, Italy \\

  E-mail: \email{lorenzo@iasf-milano.inaf.it}
}

\abstract{Supergiant Fast X-ray Transients (SFXTs) are a new class of High Mass X-ray Binaries, 
discovered by the \emph{INTEGRAL} satellite, which display flares lasting from minutes to hours, 
with peak luminosity of $10^{36} - 10^{37}$~erg~s$^{-1}$.
Outside the bright outbursts, they show a frequent long-term flaring activity 
reaching an X-ray luminosity level of $10^{33} - 10^{34}$~erg~s$^{-1}$, as recently observed with the 
\emph{Swift} satellite. Since a few persistent High Mass X-ray Binaries (HMXBs) with supergiant donors 
show flares with properties similar to those observed in SFXTs, 
it has been suggested that the flaring activity in both classes
could be produced by the same mechanism, probably the accretion of clumps composing the supergiant wind.
We have developed a new clumpy wind model for OB supergiants with both a spherical 
and a non spherical symmetry for the outflow.
We have investigated the effects of the accretion of a clumpy wind onto a neutron star in both
classes of persistent and transient HMXBs.}

\FullConference{The Extreme sky: Sampling the Universe above 10 keV - extremesky2009,\\
		October 13-17, 2009\\
		Otranto (Lecce) Italy}

\begin{document}

\section{Introduction}

In the last seven years, the hard X--ray \emph{INTEGRAL} observatory
discovered many new hard X--ray sources \cite{Bird2007}.
In particular, almost 30\% of the new discovered sources are HMXBs,
which were not detected in earlier observations.
Among these, \emph{INTEGRAL} discovered two classes of HMXBs
with supergiant companions:
the first class is composed of intrinsically highly absorbed
hard X--ray sources (e.g. IGR~J16318-4848)
\cite{Filliatre2004}.
The members of the second class, called \emph{Supergiant Fast X-ray Transients} 
(SFXTs; \cite{Sguera2005, Negueruela2006}),
exhibit outbursts with duration of a few days 
composed by many flares lasting from minutes to a few hours
as discovered by \cite{Sidoli2008, Romano2008} with
the \emph{Swift} monitoring of 4 SFXTs
(IGR~J16479$-$4514, XTE~J1739$-$302, IGR~J17544$-$2619 and AX~J1841.0$-$0536).
The behaviour of SFXTs is characterized by a high dynamic range, 
spanning 3 to 5 orders of magnitude,
from a quiescent state at $10^{32}-10^{33}$~erg~s$^{-1}$ up to
the peak luminosity during outbursts of $10^{36}-10^{37}$~erg~s$^{-1}$.
\emph{Swift} also discovered that 
SFXTs display a fainter flaring activity with luminosities of 
$10^{33}-10^{34}$~erg~s$^{-1}$.

Many different mechanisms have been suggested to
explain the SFXT behaviour:
\cite{Grebenev2007, Bozzo2008}
proposed that the high dynamic range shown by SFXTs
is due to transitions across the neutron star centrifugal barrier
produced by a change in the donor wind density.
In particular, \cite{Bozzo2008} proposed
that what distinguishes SFXTs from persistent 
HMXBs with supergiant companions is that SFXTs 
host magnetars with large spin period ($\sim 10^3$~s).
Another possibility involves the presence of an equatorial
wind component denser than the polar wind,
and inclined with respect to the orbital plane of the compact object.
In this framework, the outburst is produced when the compact object crosses the 
equatorial wind component and, consequently, accretes more matter \cite{Sidoli2007, romano09b}.
This mechanism has been successfully applied to the SFXT IGR~J11215-5952,
which shows periodic outbursts ($P_{orb}\approx 165$~days, \cite{Sidoli2007}).
\cite{intZand2005} proposed that the flaring activity in SFXTs
is due to the sudden accretion of dense blobs of matter 
composing the supergiant wind.
In the framework of the clumpy wind model proposed by \cite{Oskinova},
\cite{Negueruela2008} suggested that different orbital separations
could play a role in the different behaviour of SFXTs and persistent HMXBs.
Persistent HMXBs have a small orbital period, with a distance
supergiant-compact object $<2$ stellar radii, while in SFXTs
the compact object orbits the companion at larger distances.

\section{A new clumpy stellar wind model}

Recently we developed a new
clumpy stellar wind model for OB supergiants in HMXBs \cite{Ducci2009}.
Assuming that OB supergiants are surrounded by a clumpy and
spherically symmetric wind,
we assumed for the first time that clumps follow a power law mass distribution
\begin{equation} \label{Npunto}
p(M_{\rm cl})=k \left ( \frac{M_{\rm cl}}{M_{\rm a}} \right)^{-\zeta}
\end{equation}
where $M_{\rm cl}$ is the mass of the clump,
and  $[M_{\rm a}$ - $M_{\rm b}]$ is the mass range.
The rate of clumps produced by the supergiant is related to the
total mass loss rate $\dot{M}_{\rm tot}$ by
means of the $k$ parameter, and we defined 
$f=\dot{M}_{\rm cl} / \dot{M}_{\rm tot}$ 
as the fraction of mass lost in clumps,
where $\dot{M}_{\rm cl}$ is the component 
of mass loss rate due to the clumps.
We assumed spherical clumps,
with radii $R_{\rm cl}$, then we also
introduced a power law distribution of radii, $R_{\rm cl}$:
\begin{equation} \label{distrib R_cl}
\dot{N}_{\rm M_{\rm cl}} \propto R_{\rm cl}^{\gamma} \mbox{ \ \ clumps s}^{-1}
\end{equation}
Spectroscopic observations of O stars suggest that clumps have, on average, 
the same velocity law of a smooth stellar wind \cite{Lepine2008}. 
We can then assume for the clump velocity profile $v_{\rm cl}(r)$:
\begin{equation} \label{legge_velocita}
v_{\rm cl}(r) = v_{\infty}\left (1 - 0.9983\frac{R_{\rm OB}}{r}
\right )^{\beta}
\end{equation}
where $v_{\infty}$ is the terminal wind velocity, 
$0.9983$ ensures that $v(R_{\rm OB}) \neq 0$,
$R_{\rm OB}$ is the radius of the supergiant
and  $\beta$ is a constant
\cite{Lamers-and-Cassinelli-1999}.

From the balance pressure equation and the continuity equation \cite{Lucy-and-White-1980, Howk2000}
we find the law describing how the clump size increases with the distance 
from the supergiant star:
\begin{equation} \label{legge_Rcl_r}
R_{\rm cl}(r) = R_{\rm cl}(R_{\rm s}) \left [ \frac{r^2 v_{\rm
cl}(r)}{R_{\rm s}^2 v(R_{\rm s})} \right ]^{1/3}
\end{equation}
where $R_{\rm s}$ is the sonic radius, where the clumps start 
outflowing from the star \cite{Castor1975}.

\begin{figure}[htbp]
\begin{center}
\includegraphics[height=.4\textheight]{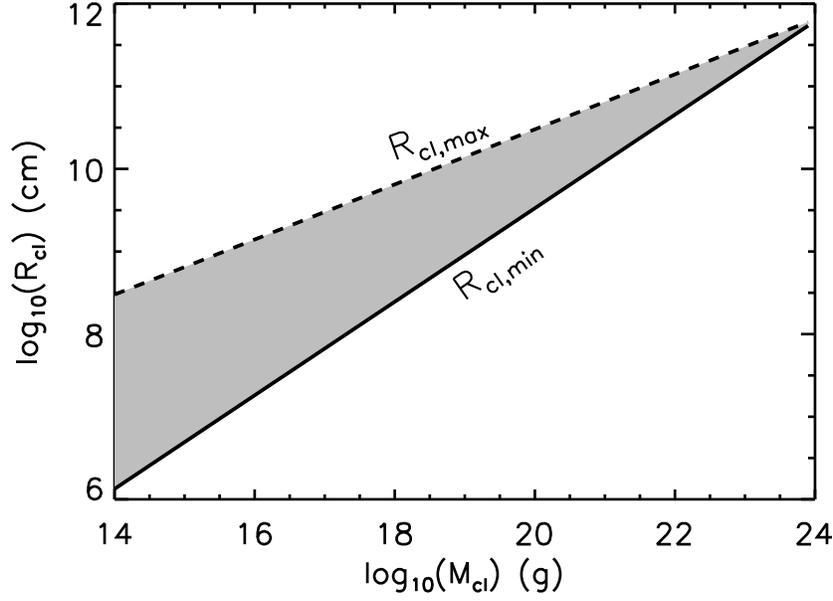}
\end{center}
\caption{Upper (dashed line) and lower-limit
  (solid line) for the clump radius at the sonic radius
  These functions have been obtained 
  assuming the following parameters: 
  $M_{\rm OB}=30$~M$_{\odot}$, $R_{\rm OB}=23.8$~R$_{\odot}$,
  $\dot{M}_{\rm tot}=10^{-6}$~M$_{\odot}$~yr$^{-1}$,
  $v_{\infty}=1700$~km~s$^{-1}$, $\beta=1$,
  $\dot{M}_{\rm cl}/\dot{M}_{\rm wind}=0.7$.}
\label{relazione_Mcl_Rcl}
\end{figure}

For any given mass of the clump, 
we derived the lower-limit for the clump radius,
starting from the assumption that, in order to be accreted
by the compact object, the clump must escape from the supergiant,
i.e. the radiative force due to the scattering 
of the ions of the clump with the UV photons must dominate 
over the gravity of the supergiant.
We also derived the upper-limit for the clump radius,
starting from the definition of the clump as a density
enhancement in the smooth stellar wind:
clumps with radii larger than the upper-limit would be
less dense than the smooth stellar wind (inter-clump medium),
in contrast with the clump definition \cite{Ducci2009}.
The upper and lower-limits for the clump radius
are drawn in Fig. \ref{relazione_Mcl_Rcl}.

To calculate the X-ray luminosity 
produced by the accretion of the inhomogeneous wind, 
we modified the Bondi-Hoyle accretion model.
Assuming different orbital configurations and clumpy wind properties,
we found that the observative characteristics of the flares
(luminosity, duration, number of flares produced), 
do not depend only on the orbital parameters, 
but potentially are also significantly affected by the properties of the clumps.
This model has been successfully applied to four HMXBs:
Vela~X-1, 4U~1700-377, IGR~J18483-0311, and IGR~J11215-5952 \cite{Ducci2009, Romano2009}.
For IGR~J11215-5952 we were able to reproduce the lightcurve observed with \emph{Swift},
with the introduction of a clumpy equatorial wind component around the supergiant.
This result is in agreement with the accretion mechanism proposed
by \cite{Sidoli2007} for this source.

\section{Comparison with the HMXB 4U~1700$-$377}
\label{Study of the SGXB 4U 1700-377 }

4U~1700$-$377 \cite{Jones-et-al.-1973} is a 
bright eclipsing X-ray binary ($P_{\rm orb}=3.412$~d)
composed by a compact object (a neutron star or a black hole),
and the O6.5~Iaf$^{+}$ star HD~153919, located at a distance of
$1.9$~kpc \cite{Ankay-et-al.-2001}.
This source is characterized by a
strong flaring activity with variations as large as a factor of
$10-100$ on short time scales (from minutes to hours) \cite{White-et-al.-1983}.

We  analyzed the IBIS/ISGRI public data archive
from 2003 March 12 to 2003 April 22, and from 2004 February 2 to
2004 March 1, for a net exposure time of $\sim 5.2$~days 
(excluding the eclipse phase). 
The data reduction was carried out using OSA~7.0, 
and from the extracted light curve ($15-60$~keV)
we found a total of $123$ flares. For each flare we
extracted the spectrum in the range $22-100$~keV. All the spectra
are well fitted by a thermal Comptonization model ({\sc comptt}
in {\sc xspec}).  
For each flare, we derived the $1-200$~keV luminosity, 
which is always greater than $5.8 \times 10^{36}$~erg~s$^{-1}$.
We then applied our clumpy wind model to the 
\emph{INTEGRAL} observations of 4U~1700$-$377.
\begin{figure*}
\begin{center}
\includegraphics[width=7.5cm]{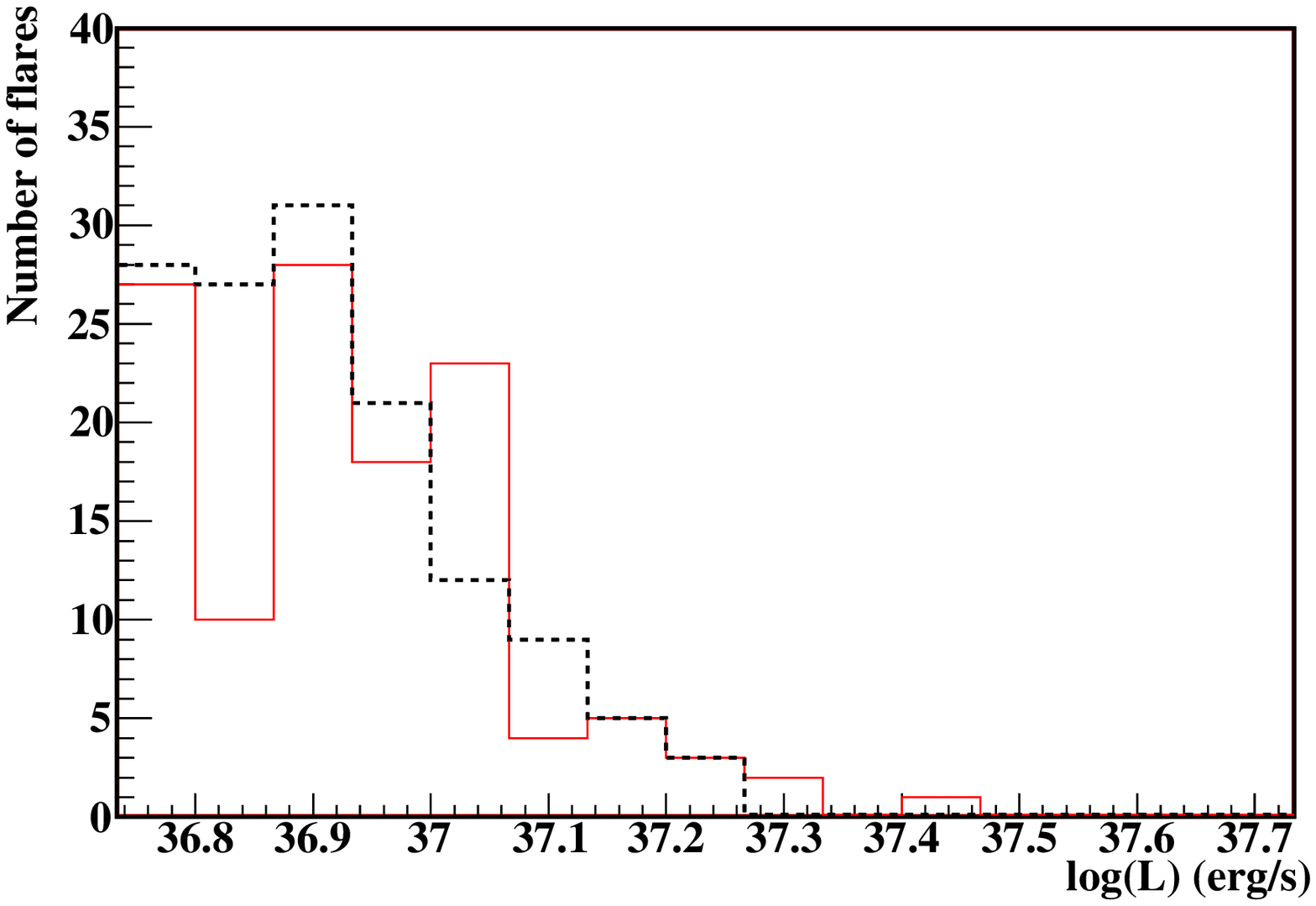}
\includegraphics[width=7.5cm]{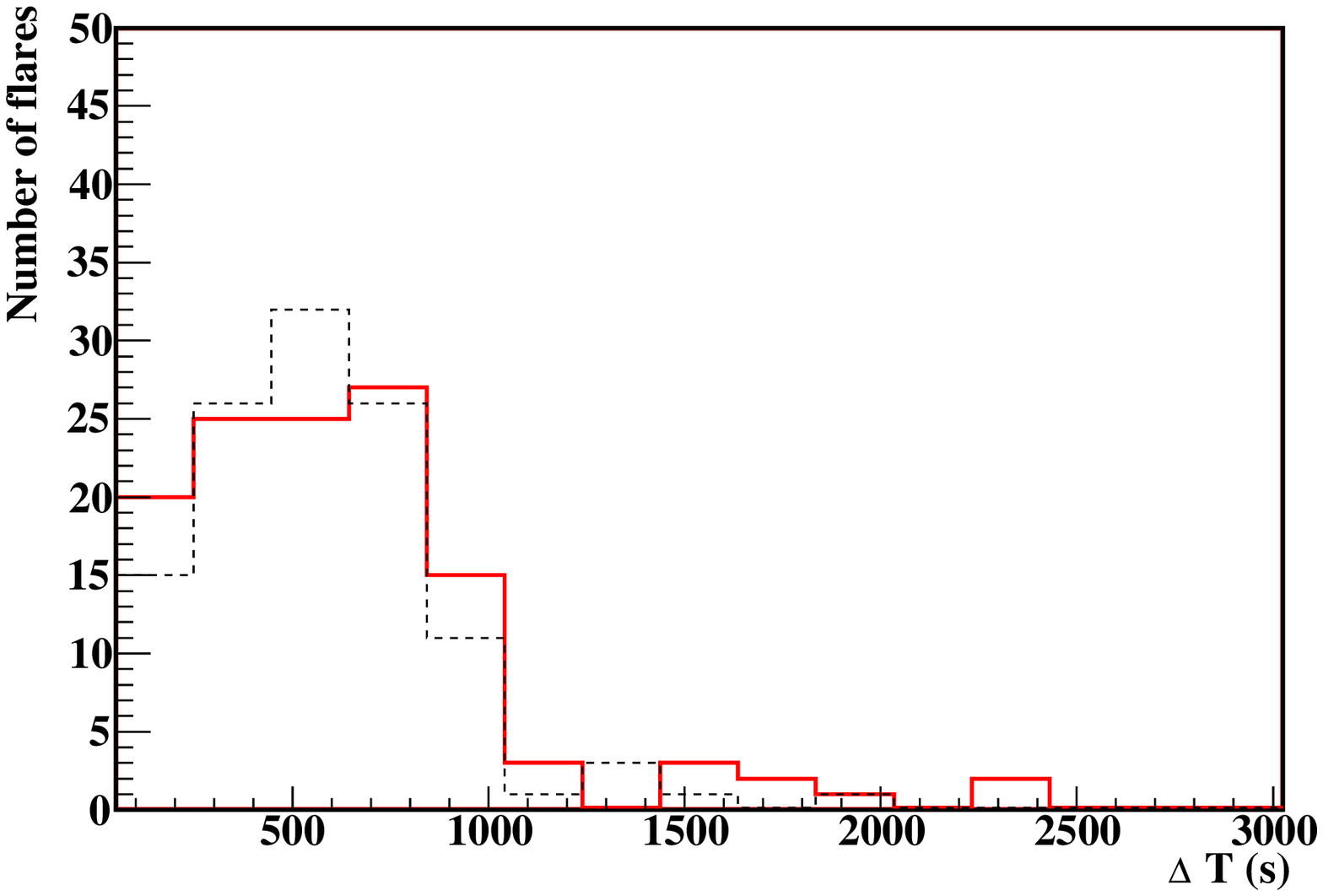}
\end{center}
\caption{Comparison of the 4U 1700$-$377 flare 
         luminosities and durations distributions,
         as observed with IBIS/ISGRI (solid line),
         with that calculated (dashed line), 
         assuming the following binary system parameters:
         $M_{\rm OB}=58$~M$_{\odot}$, $R_{\rm OB}=21.9$~R$_{\odot}$,
         $M_{\rm NS}=2.44$~M$_{\odot}$, $R_{\rm NS}=10$~km.
         The parameters for the supergiant wind are:
         $\dot{M}_{\rm tot}= 1.3 \times 10^{-6}$~M$_{\odot}$~yr$^{-1}$,
         $v_{\infty}=1700$~km~s$^{-1}$, $\beta=1.3$, $v_0=10$~km~s$^{-1}$,
         $M_{\rm a} = 5 \times 10^{16}$~g and $M_{\rm b}= 2 \times 10^{19}$~g,
         $\zeta=1.2$, $\gamma=-6.5$ and $f=0.75$.}
\label{confronto_istogrammi_4U1700}
\end{figure*}
We first compared the observed distributions of the flare
luminosities and durations with those computed adopting the system
parameters found by \cite{Clark-et-al.-2002}:
the supergiant has a luminosity $\log(L/L_{\odot})=5.82 \pm 0.07$, 
an effective temperature $T_{\rm eff} \approx 35000$~K, 
radius $R_{\rm OB} \approx 21.9$~R$_{\odot}$,
mass $M_{\rm OB} \approx 58$~M$_{\odot}$;
the mass of the compact object is $M_{\rm x}=2.44$~M$_{\odot}$.
As shown in Figure \ref{confronto_istogrammi_4U1700},
the flare properties are
well reproduced with our clumpy wind model for  $\dot{M}_{\rm
tot}= 1.3 \times 10^{-6}$~M$_{\odot}$~yr$^{-1}$, $M_{\rm a} = 5
\times 10^{16}$~g and $M_{\rm b}= 2 \times 10^{19}$~g,
$\zeta=1.2$, $\gamma=-6.5$ and $f=0.75$. 
We found that the numbers of  observed (123) and calculated flares (116)
are in good agreement.

\begin{acknowledgments}
L.D. thanks Prof. A. Treves for very helpful discussions.
This work was supported by ASI contracts  I/023/05/0,
I/088/06/0 and I/008/07/0.
\end{acknowledgments}

\end{document}